\documentclass[aip,amsmath,amssymb, twocolumn, 10pt, times]{revtex4-1}
\usepackage{graphicx,graphics}
\usepackage{dcolumn}
\usepackage{bm}
\usepackage{eqnarray,amsmath}
\usepackage{mathptmx}
\usepackage{etoolbox}
\usepackage{verbatim}
\usepackage{color}
\usepackage{ulem}
\usepackage[table]{xcolor}
\usepackage{booktabs}
\usepackage{float}
\usepackage[colorlinks=true]{hyperref}
\makeatletter
\def\@email#1#2{%
\endgroup
\patchcmd{\titleblock@produce} {\frontmatter@RRAPformat}
{\frontmatter@RRAPformat{\produce@RRAP{*#1\href{mailto:#2}{#2}}}\frontmatter@RRAPformat}
  {}{}
}%

\makeatother
\begin{document}
\preprint{AIP/JAP}
\title[]{ Channeling Skyrmions: suppressing the skyrmion Hall effect in ferrimagnetic nanostripes}

\author{R. C. Silva}
\email{rodrigo.c.silva@ufes.br}
\affiliation{Departamento de Ci\^{e}ncias Naturais, Universidade Federal do Esp\'{i}rito Santo, Rodovia Governador M\'{a}rio Covas, Km 60, S\~{a}o Mateus, ES, 29932-540, Brazil.}

\author{R. L. Silva}
\affiliation{Departamento de Ci\^{e}ncias Naturais, Universidade Federal do Esp\'{i}rito Santo, Rodovia Governador M\'{a}rio Covas, Km 60, S\~{a}o Mateus, ES, 29932-540, Brazil.}

\author{J. C. Moreira}
\affiliation{Departamento de F\'{i}sica, Universidade Federal de Vi\c cosa, Avenida Peter Henry Rolfs s/n, Vi\c cosa, MG, 36570-900, Brazil.}

\author{W. A. Moura--Melo}
\affiliation{Departamento de F\'{i}sica, Universidade Federal de Vi\c cosa, Avenida Peter Henry Rolfs s/n, Vi\c cosa, MG, 36570-900, Brazil.}

\author{A. R. Pereira}
\affiliation{Departamento de F\'{i}sica, Universidade Federal de Vi\c cosa, Avenida Peter Henry Rolfs s/n, Vi\c cosa, MG, 36570-900, Brazil.}
\date{\today}

\begin{abstract}
The Skyrmion Hall Effect (SkHE) observed in ferromagnetic (FM) and ferrimagnetic (FI) skyrmions traveling due to a spin--polarized current can be a problematic issue when it comes to technological applications. By investigating the properties of FI skyrmions in racetracks through computational simulations, we have described the nature of their movement based on the relative values of the exchange, Dzyaloshinskii--Moriya, and anisotropy coupling constants. Beyond that, using a design strategy, a magnetic channel--like nano--device is proposed in which a spin--polarized current protocol is created to successfully control the channel on which the skyrmion will travel without the adverse SkHE. Additionally, a simple adjustment in the current strength can modify the skyrmion position sideways between different parallel channels in the nanostripe.   
\end{abstract}

\maketitle
\section{Introduction and Motivation}

Skyrmions (Sks) play a unique role in low--dimensional magnets. These two-dimensional (2D) spatially localized states with broken inversion symmetry were theoretically predicted as new types of nonlinear spin textures in magnetic materials in the late 1970s~\cite{Pokrovsky, ABogdanov1989, ABogdanov}. In the last few years, they have received a great deal of effort from both fundamental science and technological applications. In a two--dimensional system, a magnetic Sk is a particle--like excitation with center magnetization pointing in the opposite direction from the border one~\cite{Everschor_2018}. Such a configuration corresponds to mapping the internal spin sphere onto the target (physical) space. The number of times such a wrapping takes place is an integer topological invariant identified with the excitation topological charge. Such topological stability prevents Sk from decaying to the ground state. Moreover, Sk--nucleation comes about due to the keen competition between exchange, anisotropy, and Dzyaloshinskii--Moriya interactions (DMI)~\cite{ABogdanov,  Dzyaloshinsky_1958, Moriya_1960, Bogdanov_1994, Robler_2006, Bogdanov_2020}. Their first direct experimental observation was reported in 2009 as Sk-crystals in ferromagnet MnSi~\cite{Muhlbauer_2009}. Thereafter, Sks have been found in a wide variety of ferromagnetic materials, even at room temperature, including bulk materials~\cite{ANeubauer2009, SXHuang_2012, Seki_2012, Karube_2016, SMuhlbauer_2016} and thin films~\cite{Yu_2010, SHeinze, Yu_2011, Tonomura_2012, XYu_2015, Moreau_Luchaire_2016, MHerve_2018, RWiesendanger, ASoumyanarayanan}. In addition,  these textures can be nucleated~\cite{Buttner_2017, SFinizio_2019, Zhang_2020, VigoCotrina_2020}, driven~\cite{WJiang_2017, KLitzius_2017, Kasai_2019},detected~\cite{AKubetzka_2017, KPalotas_2017, HZhang_2023}, and annihilated~\cite{NMathur_2021, RIshikawa_2022} by purely electronic means. These features render Sk as potential candidates for highly mobile, low--power consumption, super-dense magnetic data storage devices~\cite{AFert, Sampaio_2013, XZhang_2015, Tomasello_2014}, and skyrmion-based computation~\cite{TDohi_2023}, among other spintronic applications~\cite{NSisodia_2022, SLuo_2018, YFeng_2019, GSanchez_2016}. Nevertheless, Sk-based mechanisms in ferromagnetic (FM) systems suffer from the skyrmion Hall effect (SkHE), according to which it is unavoidably deflected from its straight trajectory due to a transverse (Magnus--like) force~\cite{NNagaosa, WJiang_2017, KLitzius_2017, Chen_2017}. Namely, SkHE deeply jeopardizes FM racetrack memory utilization once a high spin--polarized current pushes Sk to the edges where they may be trapped or even destroyed.

A promising alternative to circumvent SkHE is to depart to antiferromagnetic (AFM) systems. Actually, it has been predicted that AFM Sk's move along a straight line, suppressing the SkHE~\cite{Barker_2016, RLSilva_2019, RLSilva_2020}, although AFM SKs have been observed only in synthetic antiferromagnets~\cite{Dohi_2019, RJuge_2022, WLegrand_2020, RChen_2020}. Even in these systems, their direct observation is challenging, as current real-space imaging techniques are quite limited in dealing with textures that carry vanishing magnetization. 

Alternatively, like occurs in AFM materials, ferrimagnetic (FI) systems exhibit weak dipole fields and fast spin dynamics, yielding smaller and faster moving Sks, whenever compared to their ferromagnetic analogs~\cite{FButtner_2018, CTMa_2019}. Additionally, FI SKs have been experimentally observed in several materials, including thin films~\cite{TOgasawara_2009, MFinazzi_2013, LYe_2020, SWoo_2018, KChen_2020, ZLi_2021, YQuessab_2022, CLuo_2023} and stacked multilayers~\cite{XLiu_2009, Buttner_2017, SWoo2_2018, LCaretta_2018, RStreubel_2018, TXu_2023}. Furthermore, FI skyrmions can also be printed and deleted in a feasible and reproducible way at room temperature by using electric stimuli~\cite{Buttner_2017, SWoo2_2018}. To accomplish that, a challenge concerning its direct observation must be overcome: indeed, FM Sks have been detected by means of electrical response that SkHE leaves back~\cite{ANeubauer2009, MLee_2009, NKanazawa_2011, KHamamoto_2016}. However, for FI Sks, such effect is quite reduced, so it is the electric signal whose suitable detection is challenging for current techniques~\cite{JYu_2019}.

A FI skyrmion may be realized as composed of two Sks lying on two sublattices with different magnetic momenta, coupled by an AFM exchange term (see Section II for further details). Due to the uncompensated spins, FI Sk bears a non-zero magnetization, and it should experience a reduced SHE by virtue of the partial cancellation of the Magnus force acting on both sublattices. Consequently, it is expected to move straighter than its FM counterpart under the same stimulus.

In this work, we study FI skyrmions in stripe-like shapes framework. In the first scenario, we investigate the effects of next-nearest exchange and DM interactions on the Sk stability in a clean lattice. A number of static and dynamic quantities, such as Sk radius, Hall angle, and velocity as functions of the magnetic moment ratio of the FI sublattices are computed. Additionally, one obtains a phase diagram showing the optimal values of the second interaction coupling constants that enable stable skyrmions and how the skyrmion region is affected by modifying the difference between the spins in the distinct sublattices. In the second part, we examine a prospect in which a nano--device containing channels, i.e., narrow regions with magnetic properties different from the host material, is tailored to bypass undesirable effects such as the SkHE.

\section{Model and Methods}

In order to describe the ferrimagnetic racetrack, we consider a 2D square lattice whose magnetization is represented by an array of dimensionless vectors $\vec{\mu}_{i} = \left(\mu^{x}_{i},\mu^{y}_{i},\mu^{z}_{i}\right)$, accounting for the magnetic moment located at the lattice site \textit{i}. In this discrete system, the FI lattice can be described by two sublattices, $A$ and $B$, in which the spins of sublattice $A$ ($\mu_{A}$) have their modules larger than those located in sublattice $B$ ($\mu_{B}$). In our simulations, $\mu_{A}$ is kept fixed ($\mu_{A} \equiv 1$) while $\mu_{B}$ varies over the range $0 \leq \mu_{B}/\mu_{A} \leq 1$. Nanostripes dimensions read L$_{x}$ = 600 \textit{a}, in horizontal length, and L$_{y}$ = 150 \textit{a}, in vertical width, with \textit{a} being the lattice spacing. Thus, the strip comprises L$_{x}$L$_{y}$ magnetic moments, interacting according to:
\begin{equation} \label{eq:hamiltonian}
    \begin{array}{c}
        \mathcal{H} =  \displaystyle{\sum_{\langle i,j \rangle}}J^{AFM}_{ij}\vec{\mu}_{i} \cdot \vec{\mu}_{j} - \displaystyle{\sum_{\langle\langle i,j \rangle\rangle}}J^{FM}_{ij}\vec{\mu}_{i} \cdot \vec{\mu}_{j} -\displaystyle{\sum_{\langle i,j \rangle}}\vec{D}_{ij} \cdot \left(\vec{\mu}_{i} \times \vec{\mu}_{j}\right)\\
        \\
        -\displaystyle{\sum_{\langle \langle i,j \rangle \rangle}}\vec{D}_{ij}^{(2)} \cdot \left(\vec{\mu}_{i} \times \vec{\mu}_{j}\right)-\displaystyle{\sum_{i}k_{i}^{z}} \left(\vec{\mu}_{i} \cdot \hat{z}\right)^{2},
    \end{array}
\end{equation}
\noindent
where \textit{i} and \textit{j} are the lattice site indices. The symbol $\langle \langle ij\rangle\rangle$ in the summations indicates that we are considering next--nearest--neighbors (NN) interactions in order to introduce intra--sublattice interactions, i.e., such interaction allows magnetic moments lying in the sublattice A to interact with the others composing the same sublattice (analogously for those spins composing sublattice B). The first two terms in the Hamiltonian (Eq.~\ref{eq:hamiltonian}) are Heisenberg exchange interactions, with J$^{AFM}$ and J$^{FM}$ representing the exchange constant couplings: J$^{AFM}$ refers to AFM coupling between the first neighbor sites while J$^{FM}$ expresses a FM coupling between NN spins. The third and fourth terms account for the interfacial DM interaction. The DM vector reads $\vec{D}_{ij} = D \left(\hat{z} \times \hat{r}_{ij}\right)$, where the $\hat{z}$ is the unit vector normal to the sample's plane and  $\hat{r}_{ij}$ is the unity vector pointing from the lattice site \textit{i} to \textit{j}. This way, DM interaction supports the emergence of N\'{e}el skyrmions. $\vec{D}_{ij}^{(2)}$ represents the DM vector considering such interaction up to second neighbors. The last term in Eq.~\ref{eq:hamiltonian} is the magnetic anisotropy energy, favoring an out--of--plane orientation of the magnetic moments. k$^{z}_{ij}$ is the uniaxial anisotropy constant.\\
\begin{figure}[bt]
    \centering
    \includegraphics[width=\hsize]{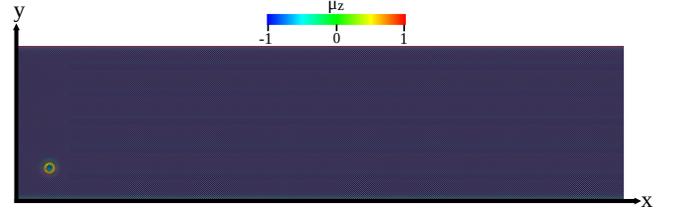}
     \caption{(Color online) Top view of the FI racetrack with the skyrmion in its initial position $\vec{r}_{sk} = (30,30)\, a$. The racetrack has length L$_{x}$ = 600 \textit{a} and width L$_{y}$ = 150 \textit{a}.}
    \label{fig:racetrack}
\end{figure}

Compact skyrmions have been recently observed in DyCo$_3$ ferrimagnet~\cite{CLuo_2023, KChen_2020, SVelez_2022}. Thus, for concreteness we use experimental parameters for such a compound. Based upon in the methodology described by V\'{e}lez {\it et al.}~\cite{SVelez_2022}, then the gyromagnetic ratio for DyCo$_3$ was calculated and a suitable time scale for it was established. Lattice spacing, $a=4$ nm, has been taken lesser than the exchange length, $\lambda_{ex}^{DyCo_3} = \sqrt{\frac{2A_{ex}}{\mu_{0}M_{s}^{2}}} \approx 4.78 \mathrm{\ nm}$. Table \ref{tab:param} summarizes the relevant parameters used in this work, with DMI and anisotropy coupling constants given in terms of $J^{AFM}$.

	\begin{table} 
	\caption{Relevant parameters along with time, space and energy scales ($\delta t$, $a$ and $J^{AFM}$) used in this work.}
	\rowcolors{2}{black!0}{black!10}
    	\centering
		\begin{tabular}{ m{50pt} >{\centering\arraybackslash} m{70pt} m{90pt}  }
		\hline
		\textit{Parameters} && \textit{In units of}  \\ \hline
		$a$   & 	$\displaystyle < \lambda_{ex}^{DyCo_3} $ & $4.0 \mathrm{\ nm}$ \\
		$\displaystyle M_s$ & $M_s^{Dy} - 3M_s^{Co}$ & $6.5 \times 10^5 \mathrm{A/m}$ \\
		$|\gamma^{eff}|$ & $\displaystyle \frac{M_s}{\frac{M_{s,Dy}}{\gamma_{Dy}} - \frac{M_{s,Co}}{\gamma_{Co}}}$ & $3.76 \times 10^7 \mathrm{rad.s^{-1}.T^{-1}}$ \\
		$J^{AFM}$ & $2aA_{ex}$ & $4.8 \times 10^{-20} \mathrm{J}$ \\
		$\delta t$	& $\frac{M_s a^3}{J^{AFM}\gamma^{eff}}\delta \tau$ &	 $2.3 \mathrm{\ ps}$		\\
		\hline
        \label{tab:param}
	\end{tabular}
\end{table}

The magnetization dynamics is determined by solving the Landau-Lifshitz-Gilbert (LLG) equation~\cite{Landau_1935, TGilbert}, augmented by the Zhang--Li spin-transfer torque term \cite{Zhang_Li}, as below:
\begin{equation} \label{eq:LLGZL}
	\begin{array}{ccc}
		\left(1+\alpha^{2}\right) \displaystyle{\frac{\partial \vec{\mu}_{i}}{\partial \tau}} = \displaystyle{-\vec{\mu}_{i} \times \vec{b}_{i} -\alpha \vec{\mu}_{i} \times \left( \vec{\mu}_{i} \times \vec{b}_{i}\right)}\\
		\\
		\displaystyle{-\nu\left\{\left(\beta - \alpha \right)\left[\vec{\mu}_{i} \times \left(\hat{j}_{e} \cdot \nabla\right)\vec{\mu}_{i}\right]+\left(1+\alpha \beta\right) \left(\hat{j}_{e} \cdot \nabla\right)\vec{\mu}_{i}\right\}},
	\end{array}
\end{equation}
\noindent
where $\alpha$ is the Gilbert damping constant, and $\vec{b}_{i} = \displaystyle{-\frac{1}{J^{AFM}}\frac{\partial \mathcal{H}}{\partial \vec{\mu}_{i}}}$ is the dimensionless local effective magnetic field acting at lattice site \textit{i}. The last two terms in Eq.~\ref{eq:LLGZL}  are the Zhang--Li spin--transfer torque, with $\nu = \frac{P j_{e} \mu_{BM} a^{2}}{|e|\gamma^{eff} J^{AFM} \left(1+\beta^{2}\right)}$. $P$ is the degree of polarization of the spin current density, $\mu_{BM}$ is the Bohr magneton, $e$ is the electronic charge, $\beta$ is the non-adiabaticity coefficient, $j_{e}$ is the magnitude of the electric current density and $\gamma^{eff}$ is the effective gyromagnetic ratio. Here, we set $\vec{j}_{e} =-j_{e} \hat{x}$ in order to drive the skyrmion from the left to the right, and $\partial_x \vec{\mu}$ is calculated in each sublattice~\cite{Barker_2016}. In our simulation, we fix $\alpha= 0.15$, $\beta=0.08$ and $P =0.90$. The LLG equation has been integrated using a fourth--order Runge--Kutta method with a dimensionless time step of $\delta \tau = 0.0001$ (the relation between real--time $\delta t$ and its dimensionless version $\delta \tau$ is given by $\delta t = \frac{M_{s}a^{3}}{\gamma^{eff} J^{AFM}}\delta \tau$, where M$_{s}$ represents the saturation magnetization). The spin--polarized current strength is given in terms of $|e|\gamma^{eff} J^{AFM}/\mu_{BM}a^{2} \approx 2.0 \times 10^{9}$ Am$^{-2}$ and the Sk velocity is measured in units of $\gamma^{eff} J^{AFM}/M_{s}a^{2} \approx 0.20 \textrm{ m/s}$. 

The initial FI Sk profile is obtained by taking an analytical solution of a single N\'{e}el Sk with radius, R$_{sk}$, placed at $\vec{r}_{sk}=\left(r_{x},r_{y}\right)$. Such a solution is described by using two scalar fields, the polar, $\theta$, and the azimuth, $\phi$, angles of the internal spin sphere, valued at each lattice site \textit{i}, with coordinates (x$_{i}$,y$_{i}$), so that: $\vec{\mu}_{i} = (-1)^{x_{i}+y_{i}} \mu_{i}\left(\sin \theta_{i} \cos \phi_{i}, \sin \theta_{i} \sin \phi_{i}, \cos \theta_{i} \right)$, where $\mu_{i}=\mu_{A}=1$, if $(-1)^{x_{i}+y_{i}}=+1$ and $\mu_{i}=\mu_{B}$ otherwise. Additionally,
\begin{equation} \label{eq:sky}
    \left\{
	\begin{array}{l}
	    \theta_{i} = \arccos{\left( \dfrac{R^{2}_{sk}-\rho^{2}_{i}}{R^{2}_{sk}+\rho^{2}_{i}}\right)} \\
	    \\
	    \phi_{i} = \arctan{\left(\dfrac{y_{i}}{x_{i}}\right)},
	\end{array}
	\right.
\end{equation}
\noindent
where $\rho_{i} = \sqrt{\left(x_{i}-r_{x}\right)^{2}+\left(y_{i}-r_{y}\right)^{2}}$. In order to enable the adjustment of Sk radius and its in--plane magnetization in the lattice, we integrate the LLG equation without the Zhang--Li term, so that the system achieves ground state, as shown in Fig.~\ref{fig:racetrack}. Such a configuration is always used as an initial condition in Eq.~(\ref{eq:LLGZL}) to investigate dynamic properties, for instance, how FI Sk moves under an applied stimulus.

Skyrmion position may be tracked back using topological charge density~\cite{CMoutafis_2009}: $\sigma_{k} = \dfrac{1}{2} \varepsilon_{ij} \vec{\eta}_{k} \cdot \left(\partial_{j} \vec{\eta}_{k} \times \partial_{i} \vec{\eta}_{k}\right)$ where \textit{i} and \textit{j} sum over the horizontal and vertical directions of the racetrack, and $\displaystyle{\vec{\eta}_{k} = \frac{\hat{\mu}_{2k}-\hat{\mu}_{2k+1}}{2}}$ is the N{\'{e}}el vector. This way, the Sk mass center is located through the position of the mean topological charge distribution, given by~\cite {LShen_2020, MStier_2021}:
\begin{equation}
    \label{eq:sky_track}
    \displaystyle{\vec{r}_{sk} = \frac{\int \vec{r} \sigma d^{2}r}{Q}},
\end{equation}
\noindent
where $\vec{r} = x\,\, \hat{x} + y \,\, \hat{y}$ is the lattice site coordinates and $Q = \int \sigma d^{2}r$ is the N{\'{e}}el topological number. Skyrmion velocity is readily obtained by $\vec{V}_{sk} =d\vec{r}_{sk}/dt = \left(V_{sk}^{x}, V_{sk}^{y}\right)$, and its radius is evaluated like below:
\begin{equation}
	\displaystyle{R = \sqrt{\int \left|\vec{r}-\vec{r}_{sk}\right|^{2}\frac{\sigma}{Q} d^{2}r}}.
 \end{equation}

\section{Numerical Results}
\subsection{Typical racetracks}
\label{sec:A}
\begin{figure}[bt]
    \centering
    \includegraphics[width=\hsize]{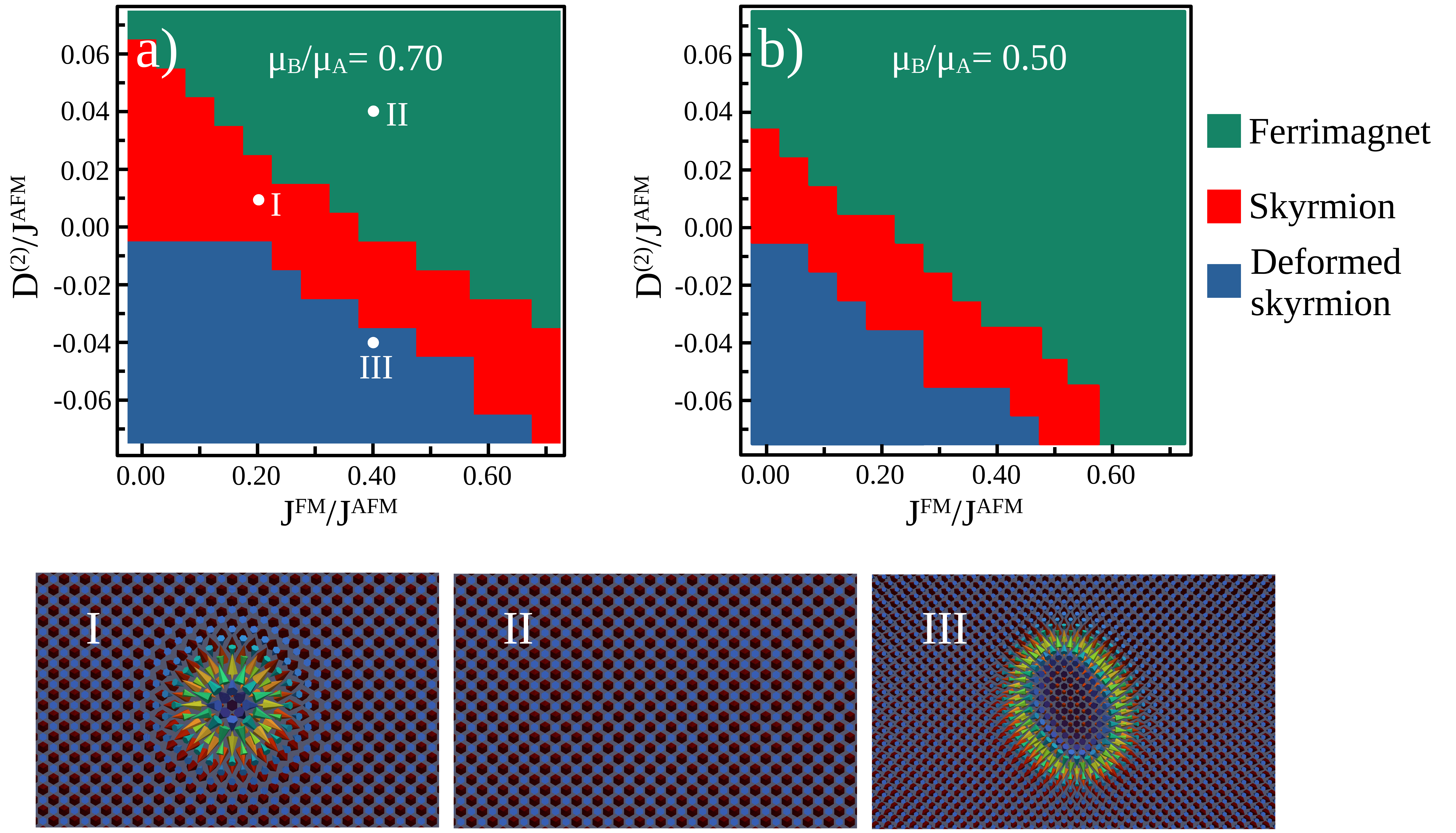}
     \caption{(Color online) Phase diagrams for FI skyrmion profile as function of J$^{FM}$ and D$^{(2)}$. Panel (a) shows the results considering $\mu_{B}/\mu_{A} = 0.70$ while in Panel (b) $\mu_{B}/\mu_{A} = 0.50$. Note that the reduction of $\mu_{B}/\mu_{A}$ ratio implies narrowing the phase space region where the FI skyrmions were found. Panels (I), (II), and (III) present typical lattice profiles in different regions of the phase diagram.}
    \label{fig:phase_diagram}
\end{figure}

We are now interested in analyzing the effects of the interaction between second neighbors in the Hamiltonian, Eq.(~\ref{eq:hamiltonian}). Disabling such interactions initially, the FI skyrmion is stabilized considering the following reduced magnetic couplings: J$^{AFM}$ = 1, D/J$^{AFM}$ = 0.20 and k$_{i}^{z}$/J$^{AFM}$ = 0.04. In the sequence, we use this initial configuration to integrate the LLG equation without the spin current terms, but including the NN interactions.
\begin{figure}[bt]
    \centering
    \includegraphics[width=\hsize]{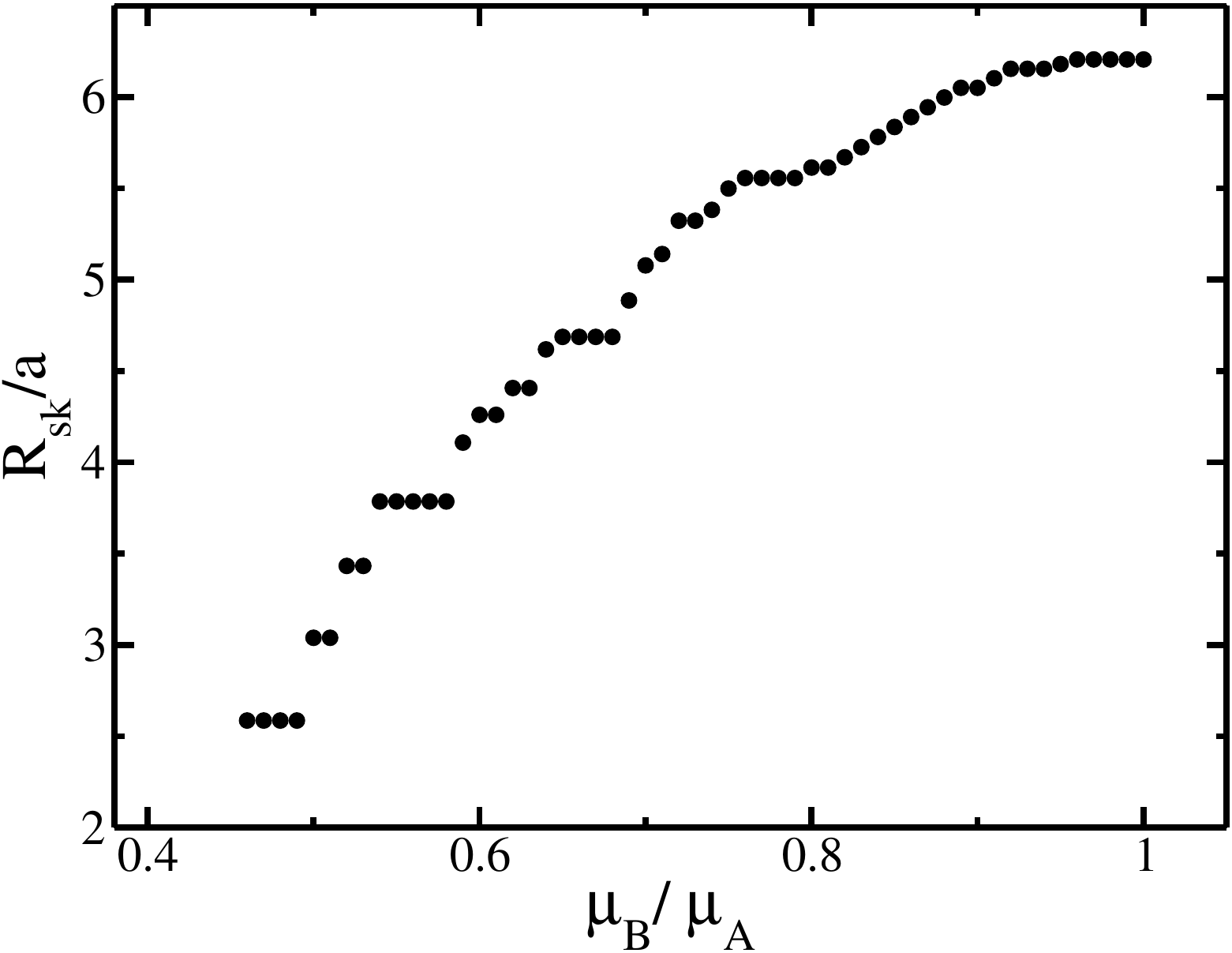}
     \caption{(Color online) Skyrmion radius, R$_{sk}$, as function of $\mu_{B}/\mu_{A}$. R$_{sk}$ monotonically decreases with the reduction of the magnetic moments lying in sublattice B. For $\mu_{B}/\mu_{A} < 0.46$, Sk is unstable and it decays to ground state.}
    \label{fig:sky_size}
\end{figure}

Varying parameters J$^{FM}$ and D$^{(2)}$ one obtains phase diagram depicted in Fig.~\ref{fig:phase_diagram}. It is noteworthy to mention its three distinct phases: FI Sk--phase (red, intermediate region), deformed Sk (blue area), and unstable  Sk's decaying to the FI ground state (green area).

The panel~\ref{fig:phase_diagram}(a) was obtained considering $\mu_{B}/\mu_{A} = 0.70$ while in panel~\ref{fig:phase_diagram}(b), one set $\mu_{B}/ \mu_{A} = 0.50$ . Clearly, the range in which skyrmions are stable is narrowing as $\mu_{B}$ decreases while the ground state is substantially increased. The bottom panels in Fig.~\ref{fig:phase_diagram} present typical magnetization profiles for different regions of the phase diagram: in (I), we have a FI skyrmion, (II) FI phase and (III) a deformed Sk phase. The NN-interactions (Exchange and DM) cause the initially circular skyrmion to deform, continually acquiring an elliptical shape. This phenomenon is commonly associated with the action of spin--polarized current~\cite{FSYasin_2022} or anti--symmetric DM interaction~\cite{YYDai_2022}. As a whole, tuning $\mu_{B}/\mu_{A}$ down yields a narrower window supporting stable FI Sk, in such a way that below a threshold no stable Sk-phase occurs, as below. Figure~\ref{fig:sky_size} shows the behavior of the skyrmion radius, R$_{sk}$, as function of $\mu_{B}/\mu_{A}$, obtained by fixing J$^{FM }$/J$^{AFM}$ = 0.10 and D$^{(2)}$/J$^{AFM}$ = 0.010. Indeed, R$_{sk}$ decreases as $\mu_{B}$ turns down. Within our framework, whenever $\mu_{B}/\mu_{A} < 0.46$, Sk size is too small and it cannot be stabilized in the discrete lattice, yielding Sk-phase suppression. So, our model supports the emergence of skyrmions only in the range $0.46 \leq \mu_{B}/\mu_{A}\leq 1.0$.
\begin{figure}[h]
    \centering
    \includegraphics[width=\hsize]{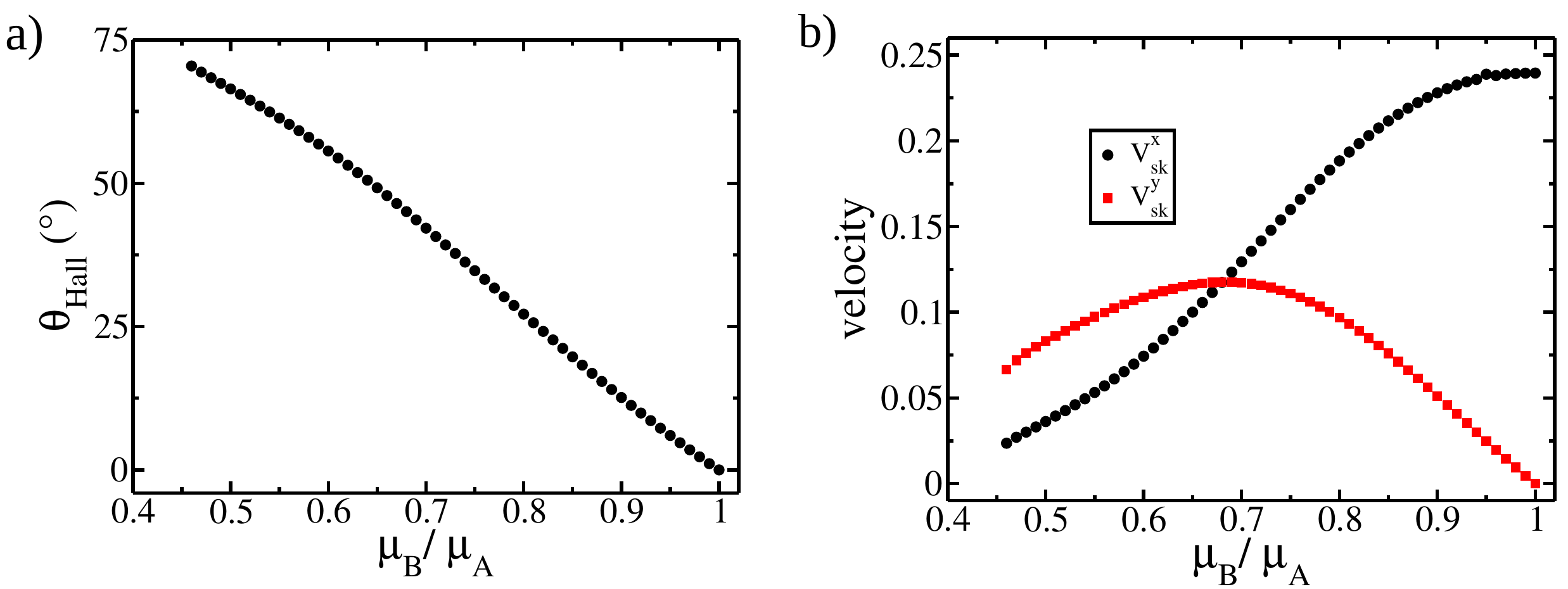}
     \caption{(Color online) a) Hall angle ($\theta_{Hall}$) and b) longitudinal (V$_{sk}^{x}$) and vertical (V$_{sk}^{y}$) velocity components as a function of $\mu_{B}/\mu_{A}$ ratio. In an AFM system, described by $\mu_{B}/\mu_{A}=1$, the skyrmion moves along the applied current and, therefore, $\theta_{Hall}$ = 0 and V$_{sk}^{y}$ = 0. Insofar as $\mu_{B}/\mu_{A}$ is reduced, $\theta_{Hall}$ and V$_{sk}^{y}$ take prominent values as the closest similarity between the system and FM media.}
    \label{fig:dyn_sky}
\end{figure}

Now, let us focus on the FI Sk dynamical properties. For the sake of comparison, let us apply the same spin--polarized current density, j$_{e}$ = 0.20, in the whole racetrack in materials with different values of $\mu_{B}$. We then analyze how SkHE depends upon $\mu_{B}$. Specifically, how the Hall angle $\theta_{Hall}$ behaves as a function of $\mu_{B}$ is shown in Fig.~\ref{fig:dyn_sky}(a). Of course, if $\mu_{B}/\mu_{A}$ = 1 (AFM case), then $\theta_{SHE}\equiv 0$, and there is no Sk deflection. Decreasing $\mu_{B}$ yields $\theta_{Hall}$ to increase almost linearly until $\mu_{B}/\mu_{A} \approx 0.60$. Facing a FI skyrmion as the overlapping of two patterns belonging to distinct sublattices, $A$ and $B$, then they carry an opposite topological charge, $\pm {\cal Q}$. However, Magnus force strength depends not only on ${\cal Q}$, but also on the sublattice magnetization, which is different in $A$ and $B$. Thus, acting on the whole excitation, forces coming from both sublattices partially cancel, resulting in a weaker SkHE acting upon a FI Sk.

How Sk velocity depends upon relative magnetization, $\mu_{B}/\mu_{A}$, is depicted in Fig.~\ref{fig:dyn_sky}(b): velocity along the applied current ($V_{sk}^{x}$, black dots) monotonically increases as $\mu_{B}$ gets higher (similar to what occurred to the Sk size), whereas the transverse component ($V_{sk}^{y}$, red dots) initially increases and reaches its maximum around $\mu_{B}/\mu_{A}=0.68$, then falls off to practically vanishes as $\mu_{B}/\mu_{A}\to 1$. Indeed, around $\mu_{B}/\mu_{A}=0.68$ both components share the same magnitude. In other words, FI Sk dynamics is quite dependent on the relative magnetization: its velocity is higher along the applied current whenever $\mu_{B}/\mu_{A}>0.68$, while for lower values, it will move faster in the perpendicular direction. Below $\mu_{B}/\mu_{A}=0.68$ both components diminish with decreasing ratio, yielding slower skyrmions when one approaches a FM system, as expected.

Using Thiele equation, Ara\'{u}jo \textit{et al.}~\cite{AAraujo_2020} have shown that Sk trajectory in a FM racetrack is given by $y(x)=(\mathcal{D}_{s}\alpha /g)x$, where $\mathcal{D}_{s}$  is the dissipative dyadic and $g=4\pi$ is associated with the gyrovector. In addition, they have also found that Sk mass reads $\mathcal{M}_{s}=4\pi L_{y}/ \alpha \gamma^{2} \sqrt{ R^{2}_{sk} +L_{y}^{2}}$, which increases as Sk radius turns down. Assuming that these results can be extended to the FI case, it is expected that the FI Sk velocity decreases according to its radius. It is also expected that the slope of the trajectory $\mathcal{D}_{s}\alpha /g \propto \mathcal{M}_{s} $ is lower in FI than in FM systems.

\subsection{SHE suppression in a designed racetrack}
\label{sec:B}
\subsubsection{Racetrack properties}

\begin{figure}[h]
    \centering
    \includegraphics[width=\hsize]{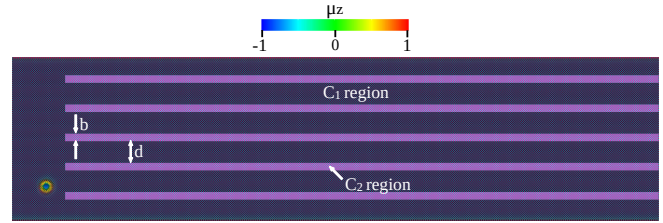}
     \caption{(Color online) Schematic top view of the proposed engineered racetrack. Such a racetrack is composed of two distinct FI materials, i.e., the magnetic properties of the C$_{2}$ region are different from those in the purple band (C$_{1}$). Sk was initially created at $\vec{r}_{sk}/a = (30,30)$, sufficiently far from C$_{2}$ region to ensure that there is no interaction between them.}
    \label{fig:eng_racetrack}
\end{figure}

In the following, we present the numerical results by solving the LLG equation with current--induced torques in an tailored racetrack, as shown in Fig.~\ref{fig:eng_racetrack}. It is composed by two distinct FI materials, i.e., the original lattice material (dark purple region) has magnetic couplings different from those of the bright purple area. Thus, the racetrack can be seen as comprised of a succession of parallel channels, each one having its own magnetic coupling. However, the ratio $\mu_{B}/\mu_{A}$ is the same for both magnetic media. The narrower region, henceforth called channel 2 (C$_{2}$), is a rectangular band of width \textit{b} and length $\ell$ = 550 \textit{a}. Between two consecutive C$_{2}$, there is a channel made from the original FI material, called channel 1 (or C$_{1}$, for short), whose width is \textit{d}. The C$_{1}$ material is characterized by the magnetic parameters: J$^{AFM}$, J$^{FM}$, D, D$^{(2)}$, k$_{z}$,  while C$_{2}$ one is described by: J$^{AFM}_{C_{2}}$, J$^{FM}_{C_{2}}$, D$_{C_{2}}$, D$_{C_{2}}^{(2)}$, k${_{z}}_{C_{2}}$. The magnetic properties of the two media support a stable Sk texture as an emergent excitation. The magnetic coupling constants at the interface between C$_{1}$ and C$_{2}$ were obtained using the geometric average: J$_{C_{1}C_{2}}^{AFM}$ = $\sqrt{\textrm{J}^{AFM}\textrm{J}^{AFM}_{C_{2}}}$, J$_{C_{1}C_{2}}^{FM}$ = $\sqrt{\textrm{J}^{FM}\textrm{J}^{FM}_{C_{2}}}$, D$_{C_{1}C_{2}}$ = $\sqrt{\textrm{D}\textrm{D}_{C_{2}}}$ and D$_{C_{1}C_{2}}^{(2)}$ = $\sqrt{\textrm{D}^{(2)}\textrm{D}_{C_{2}}^{(2)}}$. In all simulations, Sk is placed at the same initial position $\vec{r}_{0} = \left( 30 a, 30 a \right)$.

\begin{figure}[h]
    \centering
    \includegraphics[width=\hsize]{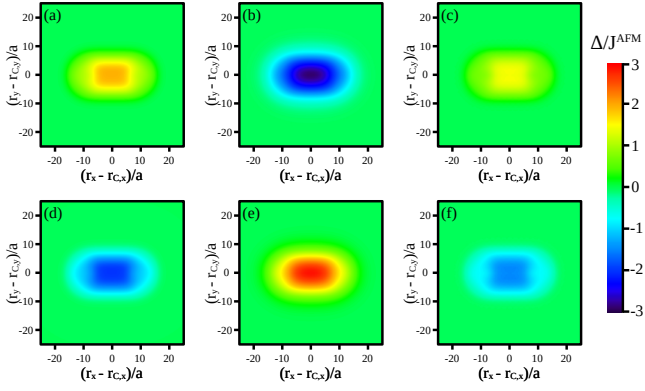}
     \caption{(Color online) Skyrmion--C$_{2}$--like defect interaction potential $\Delta$ as a function of their center of mass distance. The potentials were obtained by considering a single variation of the magnetic coupling constant of the magnetic interactions presented in Eq.~\ref{eq:hamiltonian}. In Figures a), b), and c), we consider an increase of 30$\%$ in exchange, DM, and perpendicular anisotropy coupling constants, respectively. Panels d), e), and f) were obtained by reducing the magnetic couplings of the magnetic interactions by the same value.}
    \label{fig:potentials}
\end{figure}

First, it is enlightening to understand how Sk dynamics is modified in this racetrack. The interaction potential between Sk and a single C$_{2}$ region may provide relevant information. To evaluate such an interaction, we consider a single Sk, Eq.~(\ref{eq:sky}), whose center is located at $\vec{r}_{sk} = (r_{x},r_{y})$, in the middle of the racetrack. Then, by positioning the center of C$_{2}$--like region at  $\vec{r}_{C} = (r_{C,x},r_{C,y})$, the total system's energy is calculated for this spin configuration. In fact, the length of the C$_{2}$ band on the proposed racetrack is too large, almost the same size as the length of the racetrack. In this way, it is enough to consider a C$_{2}$--like region, but with a shorter length $\ell = 10a$ and $b=6a$ to obtain the interaction potential. It gives a qualitative and straightforward comprehension of the Sk--affected dynamics. By changing the defect position $\vec{r}_{C}$, it is possible to achieve the energy as a function of the center--to--center Sk--C$_{2}$ region distance, $\vec{r}_{sk} - \vec{r}_{C}$. Such an energy is denoted as $E(\vec{r}_{sk} - \vec{r}_{C})$ and it only depends on the relative coordinate $\vec{r}_{sk} - \vec{r}_{C}$ since our simulation box is large enough to neglect any boundary effects. Note also that when the distance between Sk and C$_{2}$ is large enough,  $|\vec{r}_{sk} - \vec{r}_{C}|\to\infty$, no interaction occurs. So, we define the Sk--C$_{2}$-like interaction potential as $\Delta = E(\vec{r}_{sk} - \vec{r}_{C}) - E(\infty)$.

Figures~\ref{fig:potentials} show the potential profile, with $\mu_{B}/\mu_{A} = 0.70$ and a variation of only a single magnetic coupling of C$_{2}$ region. The other magnetic constants are the same as those used in C$_{1}$ material. Indeed, the potential brings together the contribution from the frustrated exchange, DM, and perpendicular anisotropic interactions. Therefore, this procedure enables us to determine the influence of C$_{2}$--induced changes in the contribution of each magnetic interaction considered in Eq.~\ref{eq:hamiltonian}. Figures~\ref{fig:potentials}(a) --~\ref{fig:potentials}(c) depict $\Delta$-profile considering an increase of 30$\%$ for the exchange couplings (J$^{AFM}_{C_{2}}$ = 1.30 J$^{AFM}$, J$^{FM}_{C_{2}}$ = 1.30 J$^{FM}$), DM (D$_{C_{2}}$ = 1.30 D, D$^{(2)}_{C_{2}}$ = 1.30 D$^{(2)}$), and perpendicular anisotropy (k${_{z}}_{C_{2}}$ = 1.30 k$_{z}$), respectively, concerning the couplings of the C$_{1}$ material. An increase in frustrated exchange coupling (Fig.~\ref{fig:potentials}(a)) and the uniaxial anisotropy parameter (Fig.~\ref{fig:potentials}(c)) generate a Sk--C$_{2}$ band repulsion since $\Delta > 0$. However, the same increment in the DM coupling constant (Fig.~\ref{fig:potentials}(b)) produces an opposite behavior because potential takes on negative values, indicating an attraction between the skyrmion and C$_{2}$ strip. Panels~\ref{fig:potentials}(d) --~\ref{fig:potentials}(f) represent $\Delta$ landscape where C$_{2}$ magnetic parameters are reduced by 30$\%$ with respect to those adopted in C$_{1}$ material. In this case, the decreasing in frustrated exchange (Fig.~\ref{fig:potentials}(d)) and perpendicular anisotropy (Fig.~\ref{fig:potentials}(f)) parameters yields attractive Sk--C$_{2}$ potential. In contrast, in the DM case (Fig.~\ref{fig:potentials}(e)), we have found a repulsive interaction. In all these cases, Sk--C$_{2}$ appreciable coupling occurs only if they are close enough. Indeed, $\Delta \approx 0$ for $|\left(r_{x}-r_{C,x}\right)/a|\geq 15.0$ and $|\left(r_{y}-r_{C,y}\right)/a|\geq 10$. Overall, it is possible to control Sk--C$_{2}$ coupling range by engineering the magnetic properties of C$_{2}$ material. This mechanism could be useful for SkHE suppression, as discussed below. Actually, FM and AFM Sk interacting with engineered regions have been recently reported~\cite{PLai_2017, DToscano_2019, DToscano_2020}.

\subsubsection{Skyrmion dynamics in tailored--design racetracks}

By controlling the magnetic couplings of the C$_{2}$ region, Sk dynamics can be considerably changed. Such an interaction may be attractive or repulsive, and they will be treated separately. Starting with the C$_{2}$ region repelling the Sk, we use the following values for the magnetic couplings J$^{AFM}_{C_{2}}$/J$^{AFM}$ = 1.500, J$ ^{FM}_{C_{2}}$/J$^{AFM}$ = 0.130, D$_{C_{2}}$/J$^{AFM}$ = 0.200, D$^{(2 )}_{C_{2}}$/J$^{AFM}$ = 0.013 and k$_{z_{C_{2}}}$/ J$^{AFM}$ = 0.050. We take perpendicular anisotropy smaller than that used in the C$_{1}$ material to ensure Sk stability in C$_{2}$ region.

\begin{figure}[h]
    \centering
    \includegraphics[width=\hsize]{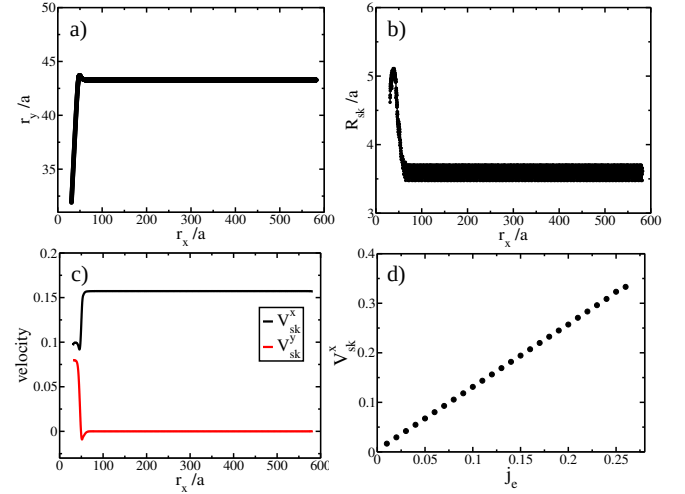}
     \caption{(Color online) a) Sk trajectory during its dynamics in the tailor-designed racetrack. b) Sk radius as a function of r$_{x}$. c) Longitudinal (V$_{sk}^{x}$) and transversal (V$_{sk}^{y}$) Sk velocity components as a function of r$_{x}$. Note that when it moves near the C$_{1}$-C$_{2}$ regions interface, the SkHE is suppressed. {d) Sk velocity as a function of spin--polarized current density j$_{e}$. Observe the direct linear relationship between $V_{sk}^{x}$  and j$_{e}$. The maximum value o j$_{e}$ that enables the SkHE suppression is 0.26. If j$_{e}$ $>$ 0.26, the Sk will cross the C$_{2}$ region, recovering its vertical movement.}}
    \label{fig:Fig7}
\end{figure}

The FI Sk is impelled to move by the action of an applied current,  $\vec{j}_{e} = -0.12 \, \hat{x}$. As expected, it exhibits a SkHE while moving in regions of the original FI material, far away from C$_{2}$ band, as shown in Fig.~\ref{fig:Fig7}(a), with $r_{x}/a < 45$, then $\theta_{Hall} \approx 38.2^{\circ}$; see also movie mv1.mp4 for further details. Henceforth, $\Delta$--potential is activated and modifies Sk further dynamics. The repulsive force between them yields Sk transversal movement (deflection). In addition, Sk size changes appreciably while displacing, as shown in Fig.~\ref{fig:Fig7}(b). The drive--force imposed by the spin--polarized current increases initially the Sk's size ($R_{sk}^{max}/a \approx 5.1$) but, as it approximates from the C$_{2}$ material interface, then Sk size is reduced, acquiring a fixed value $R_{sk}/a \approx 3.6$. In this way, while Sk approaches the region C$_{2}$, V$_{sk}^{y}$ decreases, vanishing whenever Sk starts to move in a straight horizontal line at the interface between the C$_{1}$ and C$_{2}$ bands (see Fig.~\ref{fig:Fig7}(c)). On the other hand, Sk presents a steady longitudinal velocity $V_{sk}^{x} \approx 0.16$. Fig.~\ref{fig:Fig7}(d) illustrates the linear relationship between $V_{sk}^{x}$ and j$_{e}$, consistent with Thiele analysis~\cite{SVelez_2022}. The maximum achievable velocity for the skyrmion at the interface between the C$_{1}$ and C$_{2}$ regions is $V_{sk}^{x} = 0.33$, attained when a current density $j_{e} = 0.26$ is applied throughout the racetrack. If $j_{e} > 0.26$, the SK--C$_{2}$ repulsion is insufficient to counteract the Magnus force, resulting in the Sk crossing the C$_{2}$ region.

\begin{figure}[h]
    \centering
    \includegraphics[width=\hsize]{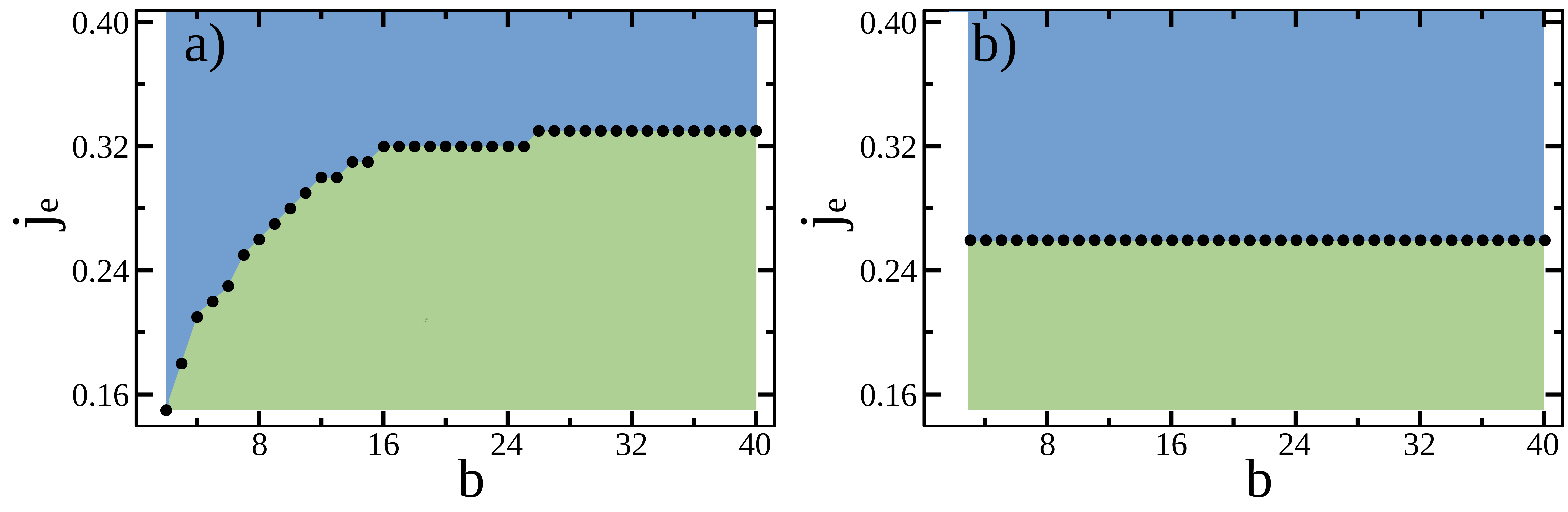}
     \caption{The spin-polarized current density j$_{e}$ as a function of the width $b$ of C$_{2}$ channel. In both graphs, the green region denotes instances where j$_{e}$ lacks the strength to propel the skyrmion past the C$_{2}$ region, resulting in movement aligned with the applied current. Conversely, the blue area represents j$_{e}$ values that facilitate the skyrmion bypassing of another vertical rail. Panel a) depicts a repulsive band, while panel b) illustrates an attractive one.}
    \label{fig:Fig9}
\end{figure}

Such a phenomenon may be understood as resulting from the competition between the Magnus force, driving Sk state, and the repulsive one, coming about due to its interaction with the channel barrier. Indeed, V{\'e}lez \textit{et al.}~\cite{SVelez_2022} showed that Magnus force acting on FI Sk--bubbles is proportional to ${j}_{e}$. In turn, the repulsion comes about due to magnetic coupling between the two channels and skyrmion--C$_{2}$ distance. Indeed, applied current must be kept below a certain threshold, $j_{e} = 0.23$, preventing Magnus force from overcoming repulsion; otherwise, Sk is eventually deflected and impelled against the borders, where it may be pinned or even annihilated. The threshold current density varies with the width b of the C$_{2}$ band, as depicted in Figs. \ref{fig:Fig9}. In Figs. \ref{fig:Fig9}(a) and (b), the green region represents j$_{e}$ values, allowing the Sk to move parallel to the applied current without SkHE. Conversely, the blue area indicates j$_{e}$ values leading to the skyrmion crossing the  C$_{2}$ channel. In Fig. \ref{fig:Fig9}(a), a repulsive C$_{2}$  region is considered, while Fig. \ref{fig:Fig9}(b) illustrates an attractive scenario.
In the repulsive case, j$_{e}$ increases with b, saturating at $j_{e} = 0.33$ for $b \geq 26 a$. Conversely, in the attractive case (see Fig. \ref{fig:Fig9}(b)), the maximum j$_{e}$ required to maintain the Sk straight-line motion remains constant ($j_{e} = 0.26$).

\begin{figure}[h]
    \centering
    \includegraphics[width=\hsize]{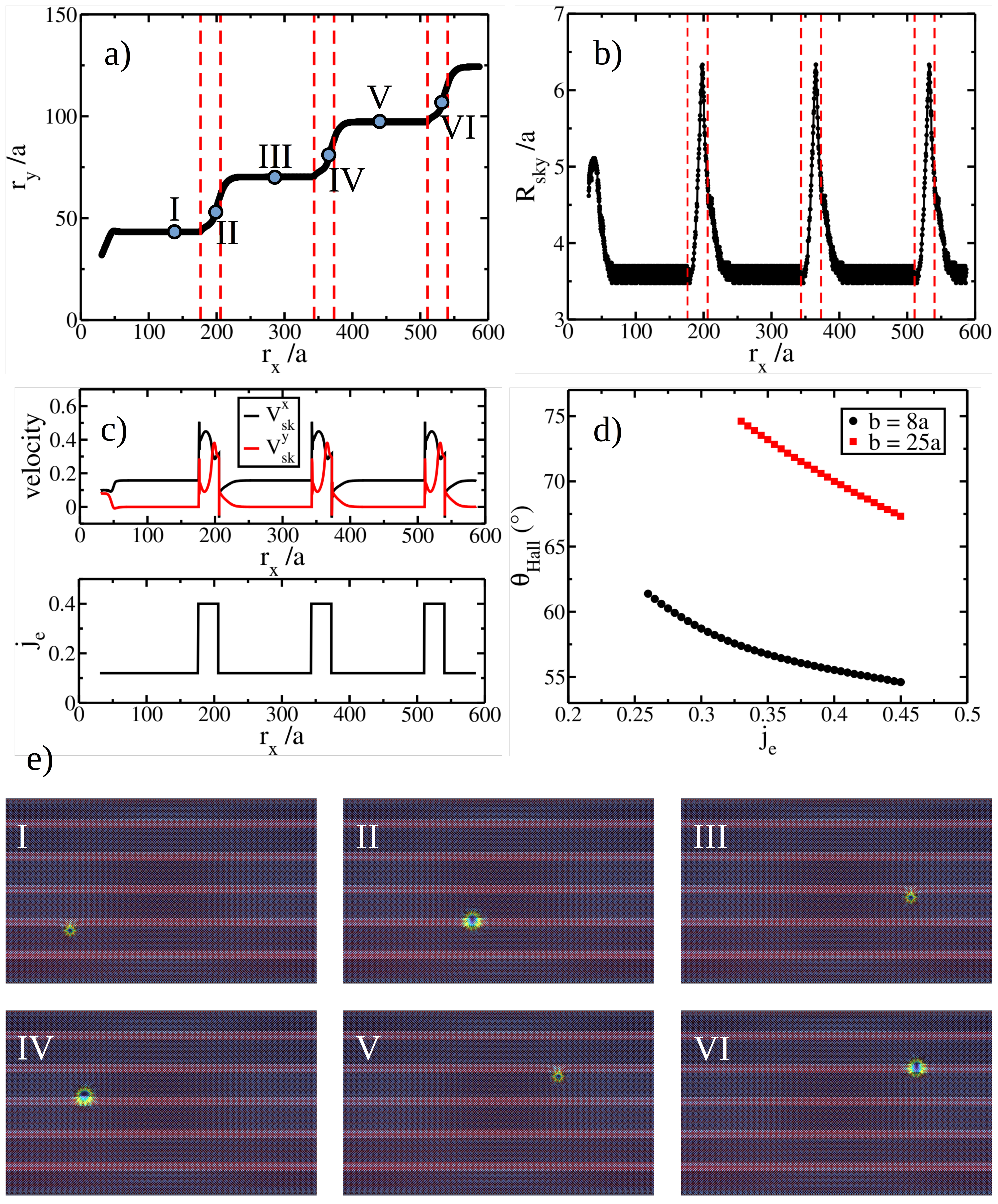}
     \caption{a) Trajectory of the skyrmion during its dynamics in the tailor-designed racetrack. b) Skyrmion radius as a function of r$_{x}$. c) Top: Longitudinal (V$_{sk}^{x}$) and transversal (V$_{sk}^{y}$) skyrmion velocity components as a function of r$_{x}$. Bottom: Current protocol is used to change the rail where the skyrmion will travel. An accuracy control of spin-polarized electric current density is mandatory to choose which rail the skyrmion will move. d) Maximum hall angle $\theta_{Hall}$ as a function of current density in racetracks with $b = 8a$ and $25a$. At lower currents, skyrmions align with current flow, causing the Hall angle to vanish. As $j_{e}$ increases, fluctuations occur, peaking notably at channel transitions. Interestingly, wider channels exhibit higher $\theta_{Hall}$ peaks, indicating a geometry--driven influence on magnetic behaviors. e) Snapshots of the FI Sk movement in the racetrack. Panels I-VI are indicated in trajectory analysis, shown in (a). Additionally, note that panels II, IV, and VI present the Sk profile while crossing C$_{2}$ band.}
    \label{fig:Fig8}
\end{figure}

A suitable current control is required to make FI skyrmion to move along a straight line. However, the possibility of crossing the energetic barrier imposed by the C$_{2}$ region through the application of electrical current above its critical value can open the doors for even a better control of skyrmions dynamical properties. One can think about creating a current protocol in which its density is increased quickly, causing Sk to travel in an adjacent rail. To investigate this possibility, we set the current at j$_{e} = 0.40$ for $\Delta \tau = 80$ (recall its initial value was j$_{e}$ = 0.12) at three different moments of Sk movement.

Figure~\ref{fig:Fig8}(a) shows Sk position while moving in the racetrack. The region between two consecutive vertical dashed red lines represents the Sk position in which the amplitude of the spin--polarized current is increased. In this case, the Magnus force wins the competition with the repulsive force, and the Sk is then driven against C$_{2}$ media, passing by the next C$_{1}$ region. The amplitude of electric current density is reduced, returning to its initial value. The Magnus and repulsive forces are balanced again, and Sk move in a straight line in the borderline that separates C$_{1}$ e C$_{2}$ bands. During the Sk passage through C$_{2}$ region, its radius increases since C$_{2}$ magnetic couplings enable its energetic stabilization with a higher size. Indeed, Fig.~\ref{fig:Fig8}(b) presents the behavior of the Sk radius, and its maximum value is R$_{sk}\approx 6.3a$, withdrawing to the previous size. The top Fig~\ref{fig:Fig8}(c) shows the longitudinal (black line) and transversal (red line) components of the Sk velocity. The longitudinal velocity component depends on the current density modulus, as predicted by the Thiele equation. The vertical one is firstly null since the Sk moves along the same direction as the applied current. During the period in which the spin current density has its magnitude increased the transverse component of the Sk velocity raises, indicating a transitory return of the SkHE. This allows Sk to overcome the C$_{2}$ channel. When the amplitude of the current density returns to its initial value,  note that the Sk still moves, displaying SkHE, as it is far from the interface of the next C$_{2}$ region. As it approaches this interface, skyrmion--interface repulsive interaction gets higher yielding to a reduced SkHE. The bottom panel in Fig.~\ref{fig:Fig8}(c) shows the current protocol employed in order to change Sk radius. Each time we increase the amplitude of the spin--polarized electric current density, the skyrmion passes through C$_{2}$ media (please, see the movie mv2.mp4 for more details). The Hall angle ($\theta_{\text{Hall}}$) of the Sk reaches its peak as the Sk transitions out of the C$_{2}$ region. In Figure~\ref{fig:Fig8}(d), these peak values are depicted with respect to $j_{e}$ for racetracks with widths of $b = 8a$ (shown as black circles) and $25a$ (depicted as red squares). Notably, the velocity of the Sk varies as it moves through different magnetic materials. Despite fluctuations in $\theta_{\text{Hall}}$ due to the unique construction of the racetrack with distinct materials, its maximum value is observed when the skyrmion transitions between channels C$_{2}$ and C$_{1}$. Furthermore, $\theta_{\text{Hall}}$ decreases with higher $j_{e}$ values and increases with the widening of the C$_{2}$ band. Additionally, Fig.~\ref{fig:Fig8}(e) presents snapshots of the Sk dynamics at various stages of its movement, highlighted in trajectory analysis (Fig~\ref{fig:Fig8}(a)). Panels I, III, and V show the skyrmion moving in the same direction as the applied current in different vertical C$_{1}$ bands. As aforementioned, in that situation, the SkHE is suppressed. Conversely, panels II, IV, and VI show the Sk crossing C$_{2}$ region due to the increase of the current density amplitude.

On the other hand, the skyrmion trajectory can be constricted in the C$_{2}$ region if we consider a magnetic material with different magnetic coupling constants. From the potential analysis, the C$_{2}$ media can attract the skyrmion if $J^{AFM}_{C_{2}} < J^{AFM}$, $D_{C_{2}}> D$ and $k_{z_{C_{2}}} < k_{z}$ (the interaction between the second neighbors follow the same behavior of the couplings in first neighbors). Here, the numerical results were obtained by considering a C$_{2}$ media with the following magnetic parameters: J$^{AFM}_{C_{2}}$/J$^{AFM}$ = 0.500, J$ ^{FM}_{C_{2}}$/J$^{AFM}$ = 0.050, D$_{C_{2}}$/J$^{AFM}$ = 0.400, D$^{(2 )}_{C_{2}}$/J$^{AFM}$ = 0.020 and k$_{z_{C_{2}}}$/ J$^{AFM}$ = 0.270. The value of the perpendicular anisotropy constant should be increased to enable the skyrmion to be a stable solution in the C$_{2}$ region. The results resemble the repulsive C$_{2}$ region presented. During the skyrmion movement, it is now attracted by the C$_{2}$ band, annulling the SkHE. The Magnus force points upward while the attractive force points downward. Then, the net force may vanish depending on the magnitude of the applied current density. Once more, we can choose which rail (C$_{2}$ material) the skyrmion will move by applying an certain spin--polarized current density. In fact, our calculations reveal some important aspects of skyrmion dynamics in this design. The skyrmion position on the racetrack can be simply manipulated. For instance, for $j_{e}< 0.24$, the skyrmion moves along a straight line, whereas for $j_{e}>0.24$, the electric current density is strong enough to pull the skyrmion out of C$_{2}$ media. Therefore, by applying a specific current protocol, the attractive C$_{2}$ region also permits the choice of which rail the skyrmion will drive (see movies mv3.mp4 and mv4.mp4 for further details). A substantial difference concerning the repulsive case lies in the geometric aspects of the racetrack: to restrict the movement of the skyrmion within region C$_{2}$, the width of this region must be $b/a >2$ and the separation between two consecutive $C_{2}$ rails must be $d /a > 7$. Otherwise, the skyrmion is annihilated while traveling on the racetrack. Moreover, in this case, Sk velocity is lower than in the scenario where it travels in C$_{1}$ media.

Actually, proposals for channeling skyrmions and other textures have appeared in several recent articles. For instance, in the work of Reichhardt {\it et al.}~\cite{CReichhardt_2016}, Magnus force is used to drive skyrmions on a quasi--one--dimensional system including the possibility of speed up effect coming about by periodic pinning potential, while a quantized SK--transport is predicted to occur in a two--dimensional periodic substrate~\cite{CReichhardt_2015}. In turn, solitonic textures are verified to present a controlled motion while moving in skyrmion chains, as reported in Refs.~\cite{NPVizarim_2022, JCBSouza_2023}. Additionally, diode--like effects have been investigated in channels with periodic potentials~\cite{JCBSouza_2022} and funnel--type geometries~\cite{JCBSouza_NJP_2022}. Numerous effects attributed to pinning defects on ferromagnetic skyrmions and other magnetic textures in one- and two-dimensional systems is reviewed in the work of Ref.~\cite{CReichhardt_2022}. In a forthcoming paper we intend to address how point-like and finite-size pinning defects along with temperature affects ferrimagnetic skyrmion structure and dynamics~\cite{JCMoreira}.

Indeed, the drive and control of magnetic textures, like skyrmions, vortices, domain walls etc, can greatly benefit whenever such patterns are considered along with artificial spin ice arrangements. Nowadays, such systems can be easily built with any underlying geometry, including usual (triangular, square, rectangular etc)~\cite{LASMol_2009, LASMol_2012} and exotic ones (brickwork, pinwheel etc)~\cite{SHSkjaevo_2020, PSchiffer_2021}. The main point here is that once spin ices bear geometric frustration, their behavior is quite distinct from usual interacting magnets. In this framework, energetically favorable configurations may work as efficient stimuli to drive magnetization patterns, like skyrmions, making them faster and controlled in magnetic assembled media.

\section{Summary}

We have investigated the static and dynamic behavior of a single skyrmion in ferrimagnetic nanostrips. Considering the Heisenberg exchange and DM interactions until the second neighbor interaction, we have obtained a narrow region of the phase diagram where stable skyrmions are observed. On the other hand, some relevant skyrmion properties are also strongly affected by modification in the $\mu_{B}/\mu_{A}$ ratio as its size, velocity, and Hall angle. Based on these results, we have suggested an example using nanosized ferrimagnetic channels, tailoring the energy landscape of nanostrips in such a way that the SkHE is suppressed. This approach can also be utilized to engineer the FI skyrmion movement, making it travels from one lane to another simply by controlling the magnitude of the applied spin-polarized current. The current power would work like a key, adjusting the skyrmion transversal position on demand along the nanostripe in a prospective device. Such a switch must be very relevant for new technologies like skyrmionics in which a complete manipulation of the skyrmion trajectories is of fundamental importance.        

\vskip .3cm
{\bf Acknowledgements\\}
The authors thank financial support from CAPES, CNPq, FAPEMIG, FAPES, INCT/CNPq - {\it Spintr\^onica e Nanoestruturas Magn\'eticas Avan\c{c}adas (INCT-SpinNanoMag)}, and {\it Rede Mineira de Nanomagnetismo/FAPEMIG}.

\section*{Data Availability Statement}
The data that support the findings of this study are available from the corresponding author upon reasonable request.

\bibliography{ferrimagnetic}

\end{document}